\documentstyle[epsfig,aps,prb,multicol]{revtex}
\begin{document}

\title{Fractional charge revealed in computer simulations of resonant
tunneling in the fractional quantum Hall regime}

\author{E.V. Tsiper}

\address{
 School of Computational Sciences, George Mason University, Fairfax,
VA 22030\\
 Center for Computational Materials Science, Naval Research
Laboratory, Washington, DC 20375
 \\etsiper@gmu.edu
}

\date{May 2, 2006}

\maketitle

 \begin{abstract}
 The concept of fractional charge is central to the theory of the
fractional quantum Hall effect (FQHE).  Here I use exact
diagonalization as well as configuration space renormalization (CSR)
to study finite clusters which are large enough to contain two
independent edges.  I analyze the conditions of resonant tunneling
between the two edges.  The ``computer experiment'' reveals a periodic
sequence of resonant tunneling events consistent with the
experimentally observed fractional quantization of electric charge in
units of $e/3$ and $e/5$.
 \end{abstract}

{\ }
\vskip -3 in
\noindent
\hfill accepted to Phys. Rev. Lett. (2006)
\vskip 2.9 in

\begin{multicols}{2}

Perhaps, the most intriguing feature of the FQHE\cite{tsui} is the
existence of quasiparticles whose electric charge is a simple fraction
of the elementary charge $e$.\cite{laughlin} Quasiparticles of charge
$e^*=e/3$ and $e/5$ have been first observed experimentally in the
$\nu=\frac{1}{3}$ and $\nu=\frac{2}{5}$ fractional states,
respectively, using resonant tunneling via a quantum antidot (a
potential hill).\cite{goldman-su,goldman1,goldman2}

Since the bulk fractional state is an insulator, the most interesting
transport properties of the system are associated with the edges,
particularly with tunneling into or between the
edges.\cite{goldman-su,goldman1,goldman2,chang,boyarsky,khlebnikov}
Due to cluster size limitations, computational studies of edge physics
have focused on the properties of a single edge, such as
non-universality of the tunneling exponent\cite{rezayi2005,goldman} or
reconstruction of the charge
density.\cite{mitra,rezayi2002,stripes,jain2005} The properties of the
Laughlin wave-function describing a dual-edge system have been studied
in cylindrical\cite{haldane} and disk geometries.\cite{yang} Study of
edge to edge tunneling through a bulk fractional state requires
clusters large enough to contain two independent edges.

Exact diagonalization (ED) of finite clusters has been very fruitful
in helping to understand the physics of
FQHE.\cite{laughlin,rezayi2005,goldman,mitra,rezayi2002,stripes,jain2005,haldane,yang}
Ordinary electronic structure methods fail for this system because the
kinetic energy is quantized by the magnetic field.  ED, or ``Full CI''
in quantum chemical terminology, imposes severe restrictions on the
cluster size, since the dimensionality of the Hilbert space grows
exponentially with the number of particles $N$:

\begin{equation}
\left(\begin{array}{c}L\\N\end{array}\right)
 \approx\frac{1}{\sqrt{2\pi Lf(1-f)}}\left[\frac{1}{f^f(1-f)^{(1-f)}}\right]^L
 \label{Nb}
\end{equation}
 Here $L=N/f$, $f<1$ is the {\em filling factor}.  The quantity in
brackets reaches maximum value of 2 at $f$=$\frac{1}{2}$.

In the past we have perfected the Lanczos
technique,\cite{lanczos,parlett} both Hermitian\cite{2D,1D} and
not,\cite{lanc,oblique,tdhf} to work with matrices up to
$\sim10^9\times10^9$.  Equation (\ref{Nb}) translates this into about
$N$=12 particles at $f$=$\frac{1}{3}$ or $N$=16 at $f$=$\frac{1}{2}$.
Whereas exact solutions for up to $N$=22 are sometimes
possible,\cite{1D} $N\gtrsim12$ normally require approximate methods.
Here I use ED and also an approximate method to model the resonant
tunneling experiments.\cite{goldman-su,goldman1,goldman2}

In\cite{goldman-su,goldman1,goldman2} a periodic sequence of {\em
resonant tunneling events} was observed as either the magnetic field
$H$ or the backgate voltage $V_{\rm BG}$ were varied.  The tunneling
events are thought of in terms of a quasiparticle tunneling through the
bulk of the fractional state between the outer edge of the sample and
the inner edge formed around the antidot.  The periodicities $\Delta
H$ and $\Delta V_{\rm BG}$ were related\cite{goldman-su} to the
quasiparticle charge $e^*$.

 \centerline{\epsfig{file=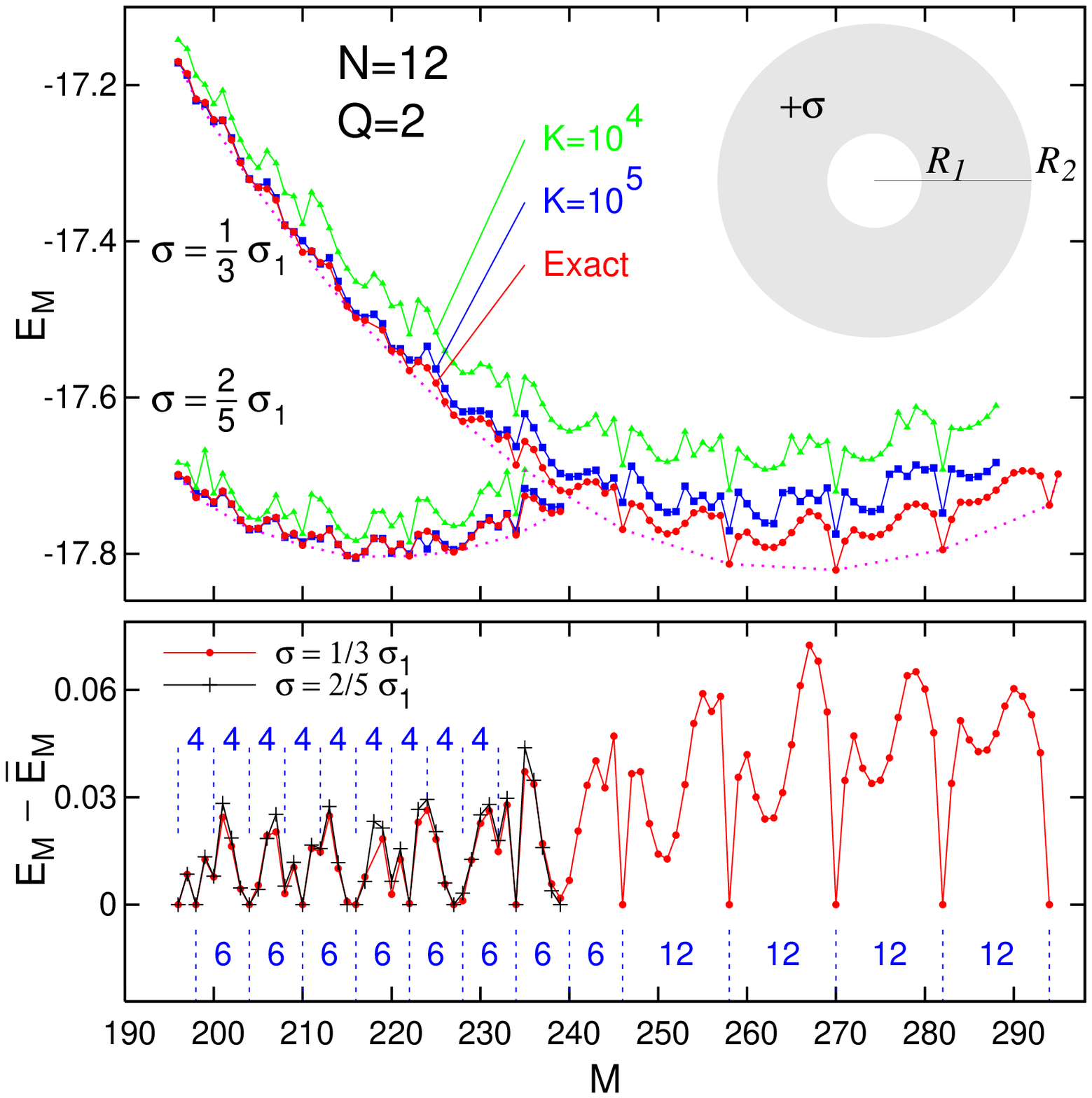,width=8.9cm}}


 {\small {\bf Fig.~1} Exact ground-state energy of $N=12$ electrons in
2D confined by Coulomb attraction to a uniformly-charged annulus
(inset) of charge density $\sigma=\frac{1}{3}$ and $\frac{2}{5}$ of
the filled Landau level.  The units are $e^2/\ell_H$.  The upper panel
also shows CSR results for $K=10^4$ and $10^5$.  Full Hilbert space
dimensionality is $10^8-10^9$ and varies with $M$.\cite{comment} The
$\frac{2}{5}$ data is shifted upward by 1.5.  The lower panel exposes
the FQHE-related structure in $E_M$ by subtracting the greatest convex
minorant $\overline E_M$ [the dotted purple line].}

 \vskip 0.1 in

In order to mimic the experimental setup I consider a planar FQHE
sample with two unconnected edges (inset in Fig.~1).  $N$ electrons in
the lowest Landau level are confined by the potential of a
uniformly-charged disk with a hole in the center, positioned in the
plane of the two-dimensional (2D) electron gas.  The positive charge
density $\sigma$ and the inner radius $R_1$ of the disk are free
parameters.  The outer radius $R_2$ is always chosen such that the
whole system is neutral.  The electronic density $\rho(r)$ confines
itself between $R_1$ and $R_2$, falling off sharply beyond this range.
Setting $\sigma$ to a fraction $\nu=\frac{1}{3}$, $\frac{2}{5}$,
etc.~of the density $\sigma_1$ of the completely filled Landau level
controls the fractional state, with $\rho(r)$ approaching
$\nu\sigma_1$ (for $N\rightarrow\infty$) far from both edges. 
 Near the edges $\rho(r)$ is known to exhibit oscillatory behavior
thought to decay slowly into the bulk.\cite{stripes}
 Such behavior prevents formation of a well-defined density plateau
between the edges in the finite clusters studied here numerically.

An increase in $R_1$ strengthens the antidot and expels charge from
inside of the antidot towards the outer edge.  The charge expelled
does not accumulate in the bulk because of neutrality considerations
and because of incompressibility of the bulk fractional state.  I
prefer to use the ``missing charge'' $Q=\sigma\pi R_1^2/e$ as a
variable, instead of $R_1$.  When $Q$ is continuously increased, the
ground state of the system reconstructs via a step-like process.  The
reconstruction events correspond to ground state degeneracies, when it
costs no energy to transfer charge from the inner to the outer edge.
This is precisely the condition for resonant tunneling through the
antidot.

In the disk geometry the single-particle states $\psi_m$ in the lowest
Landau level are characterized by the angular momentum $m=0,1,...$,
and the total angular momentum $M=\sum m$ is conserved.  The Coulomb
matrix elements are known.\cite{girvin,elm} The matrix elements of the
confining potential are $V_m=V_m(R_2)-V_m(R_1)$, where

 \begin{eqnarray}
 &&V_m(R)=\int\int_{\rho<R}d^2\rho\ d^2r
  \frac{e\sigma}{|{\bf r}-\bbox{\rho}|}|\psi_m(r)|^2\\
 &=&(2\pi)^{3/2}e\sigma\ell_H Ze^{-Z}
  \sum_{i=0}^m[q^-_{im}I_0(Z)+q^+_{im}I_1(Z)]Z^i.
  \nonumber
 \end{eqnarray}


 Here $\ell_H=\sqrt{\hbar c/eH}$ is the magnetic length,
$Z=R^2/4\ell_H^2$, $q^\pm_{00}=1$, $q^\pm_{0m}=(2m\pm1)q_{0,m-1}$,
$q^\pm_{im}=(2m-2i\pm1)q^\pm_{i,m-1}+2q^\pm_{i-1,m-1}-2q^\mp_{i-1,m-1}$.

For a given set of $N$, $Q$, and $\sigma$ I find the lowest energy
$E_M(Q)$ at each $M$.  The ground state energy is then $E(Q)=\min
E_M(Q)$.  The ground state reconstruction events occur via level
crossings of branches with different $M$ and lead to a step-wise
function $M(Q)$.  The number $p$ of the steps that occur per $\Delta
Q=1$ may be related to the charge $e/p$ that is moved from the inner
to the outer edge per one reconstruction event.

Figure 1 shows $E_M(Q)$ for $N=12$, $Q=2$. The sequence of sharp cusps
on the right curve are the $\frac{1}{3}$ fractional states.  The state
at $M=270$ is the true ground state at $Q=2$ and
$\sigma=\frac{1}{3}\sigma_1$, whereas the states with $M=270\pm12$,
$M=270\pm24$, etc.~are candidates for the ground state at different
$Q$.  The quasi-periodicity with $\Delta M=12$ is due to the
approximate invariance of the antidot Hamiltonian with respect to the
Laughlin's quasihole creation operator\cite{laughlin} ${\cal A}_0$,
also known to be the generator of infinitesimal magnetic
translations.\cite{rashba} Applied to an arbitrary many-electron wave
function $\Psi$, it translates it in the angular momentum space,
$m\rightarrow m+1$.  The total angular momentum then transforms as
$M\rightarrow M+N$:

 \begin{equation}
 \Psi_{M+N}\approx{\cal A}_0\Psi_M
 \label{approx}
 \end{equation}
 At $Q=0$ the ground state occurs, approximately, at the Laughlin's
angular momentum\cite{yannouleas}

 \begin{equation}
 M^*(Q=0)=\frac{N(N-1)}{2\nu}.
 \label{mstarL}
 \end{equation}
 I generalize this for a disk with the missing charge $Q$ as

 \begin{equation}
 M^*=\frac{(N+Q)(N+Q-1)}{2\nu}-\frac{Q(Q-1)}{2\nu}.
 \label{mstarr}
 \end{equation}
 For $N=12$, $Q=2$ I get $M^*_{1/3}=270$ and $M^*_{2/5}=225$
(cf. Fig.~1).

The single-particle orbitals $\psi_m(r)$ are localized near
$r=\ell_H\sqrt{2m}$, where $\ell_H$ is the magnetic length.
Therefore, in a macroscopic system whose density approaches a constant
$\nu\sigma_1$ in the bulk, the operator ${\cal A}_0$ pushes the
density out of the center, creating an effective positive charge
$e^*=\nu e$.  This is an exact formal property of ${\cal A}_0$ but
relates to the physical system via the approximate invariance
(\ref{approx}).  Indeed, Eq.~(\ref{mstarr}) can be obtained from
(\ref{mstarL}) by applying ${\cal A}_0^{(Q/\nu)}$:

\begin{equation}
M^*=M^*(Q=0)+NQ/\nu.
\label{mstarr1}
\end{equation}
 Figure 1, therefore, suggests that the ground state of the
$\frac{1}{3}$ system changes in steps of $\Delta M=N$, transferring
charge $e^*=e/3$ from the inner to the outer edge at every step.
According to (\ref{mstarr1}) the steps should occur at $\Delta
Q\approx\nu=\frac{1}{3}$.

Remarkably, the range of $M$ that corresponds to the $\frac{2}{5}$
fractional state exhibits double periodicity, $\Delta M=6=N/2$.  This
should cause branch crossings twice as often, $\Delta
Q\approx\nu/2=1/5$.  Consequently, the charge transferred per one
reconstruction event is $e^*=\nu e/2=e/5$, in agreement with the
experiment.\cite{goldman1} The double periodicity in $E_M(Q)$ at
$\sigma=\frac{2}{5}\sigma_1$ shows up also for $N=11$, 10, 9, and 8,
though it is less pronounced for smaller $N$.

Figure 1(b) exposes the tiny structure in $E_M(Q)$ by subtracting its
greatest convex minorant $\overline E_M$ [dotted purple line in
Fig.~1(a)].  I notice that this structure, which is the manifestation
of the FQHE, is insensitive to the confining potential, and is
practically the same for $\sigma=\frac{1}{3}\sigma_1$ and
$\frac{2}{5}\sigma_1$, suggesting rigidity of the wave function with
respect to the confining potential [$\delta\Psi_M/\delta V(r)$ being
small by some measure], whereas the principal effect of $V(r)$ is to
select which $M$ is the ground state.  Figure 1(b) also suggests that
traces of $\Delta M=4$ periodicity may be present in the area near
$M^*_{3/7}=210$.  By the same logic, $\Delta M=N/3$ at $\frac{3}{7}$
filling corresponds to $e^*=e/7$, although larger clusters are needed
to separate the $\frac{2}{5}$ and $\frac{3}{7}$ states.  Indeed, from
Eq.~(\ref{mstarL}), the condition $M^*_{2/5}-M^*_{3/7}\geq N$ leads,
at $Q=0$, to $N\geq 13$.

I further use the CSR approach to study larger clusters up to $N$=15
and to compute $E_M(Q)$ for several values of $Q$ and obtain the
branch crossings directly.  The CSR reduces the dimensionality of the
many-body Hilbert space by selecting its most relevant subspace.  It
is similar in spirit to the basis set reduction (``BSR'') technique by
Wenzel and Wilson\cite{wilson} and a number of related
algorithms.\cite{knowles,dagotto,efros,modine} The CSR iteratively
improves a basis set of $K$ many-body configurations using their
weight in the solution as the criterion of relevance.

The CSR algorithm works as follows.  I start with an arbitrary set of
$K$ many-body configurations and diagonalize the Hamiltonian in the
subspace they span.  The result is a state vector expanded in
many-body configurations.  I then retain $K'<K$ configurations by
discarding the ones which have little weight.  I re-expand the set
with the new configurations that have large matrix elements with those
retained.  When repeated, the procedure converges after 10---15
iterations (Fig.~2) and yields some optimal set of
configurations.

 \centerline{\epsfig{file=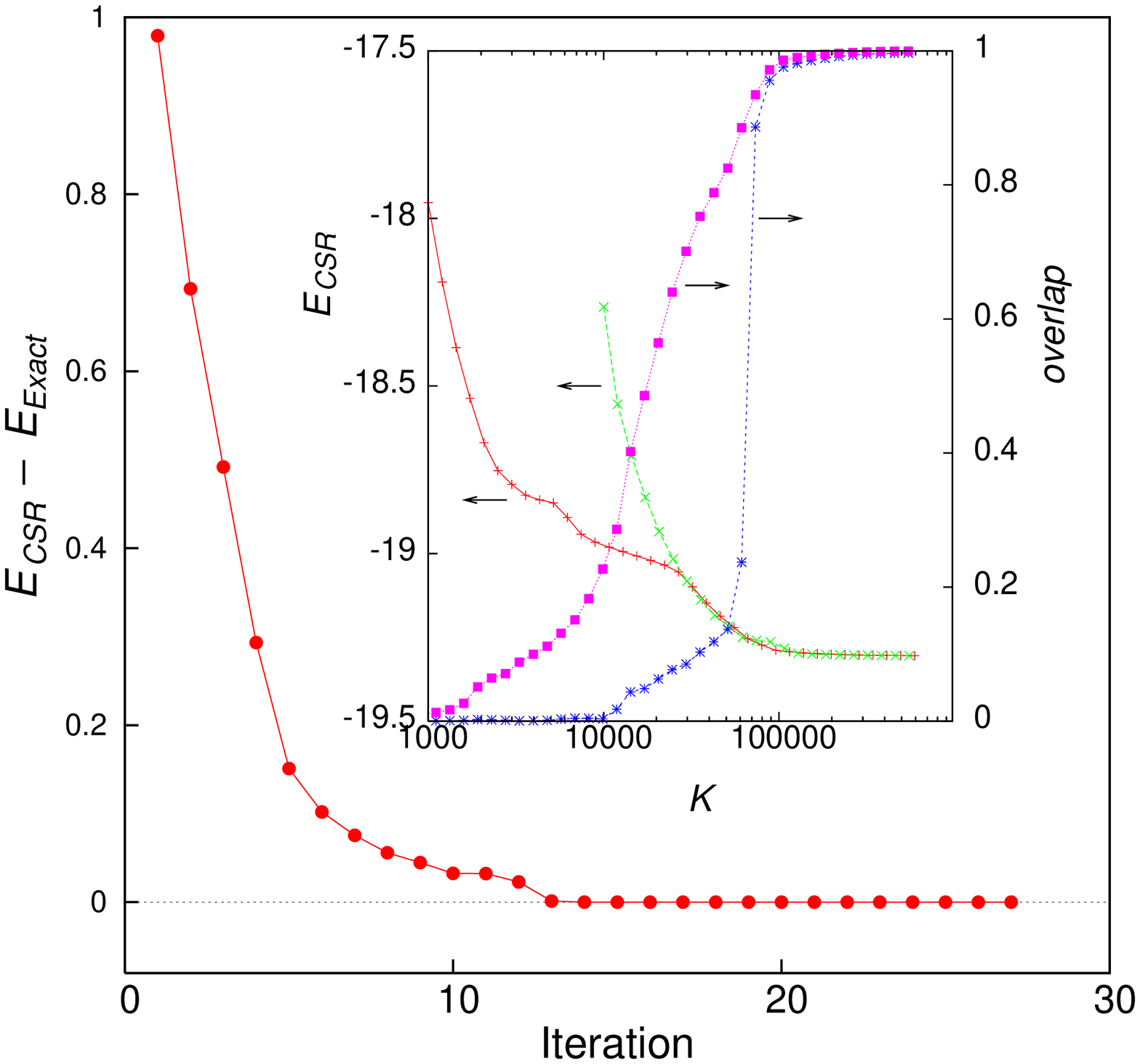,angle=-90,width=8.9cm}}

 {\small {\bf Fig.~2} Typical CSR energy convergence, $N=12$,
$\sigma=\frac{2}{5}\sigma_1$, $M=216$, $K=10^5$.  The inset shows
convergence with $K$ when $K$ is increased by a factor 1.2 at every
iteration.  ``+'' (red) and ``$\times$'' (green) data sets indicate
insignificance of the starting value of $K$ except for the initial
iterations.  The stars (blue) give overlap with the exact eigenvector.
The squares (purple) show the magnitude of projection of the exact
eigenvector onto the CSR subspace (that is, maximum overlap with any
vector in the subspace), and as such characterize the quality of the
subspace.\cite{comment1}}

 \vskip 0.1 in

The resulting basis truncation is essentially many-body, and cannot be
achieved by truncating or rotating the single-particle basis.  ED
performed in the subspace yields a variationally-stable ground state
energy that converges to the exact value as $K$ is increased.  In
practice, I do not keep $K$ constant, but increase it from iteration
to iteration (Fig.~2, inset), monitor the convergence, and extrapolate
as $1/K\rightarrow 0$.\cite{comment0} 

Figure 1 compares CSR results for $K=10^4$ and $10^5$ against the
exact solution.  We see that the essential FQHE structure survives the
basis truncations of several orders of magnitude, and that a
reasonable accuracy is achieved for $K=10^5$ in the whole range of
$M$.  Qualitative results are obtained already with $K$ as small as
$10^4$.

I used CSR to compute $E_M(Q)$ for $Q$ in steps of 0.25, interpolated
between these points with a cubic spline and found $\min E_M(Q)$ over
$M$ at every $Q$ (Fig.~3).  I used $K=200,000$ for $N\leq13$,
$K=500,000$ for $N=14$, and $K=1,000,000$ for $N=15$.  $K$ was doubled
in some calculations where the extrapolation to $1/K\rightarrow0$
seemed ambiguous.

 \centerline{\epsfig{file=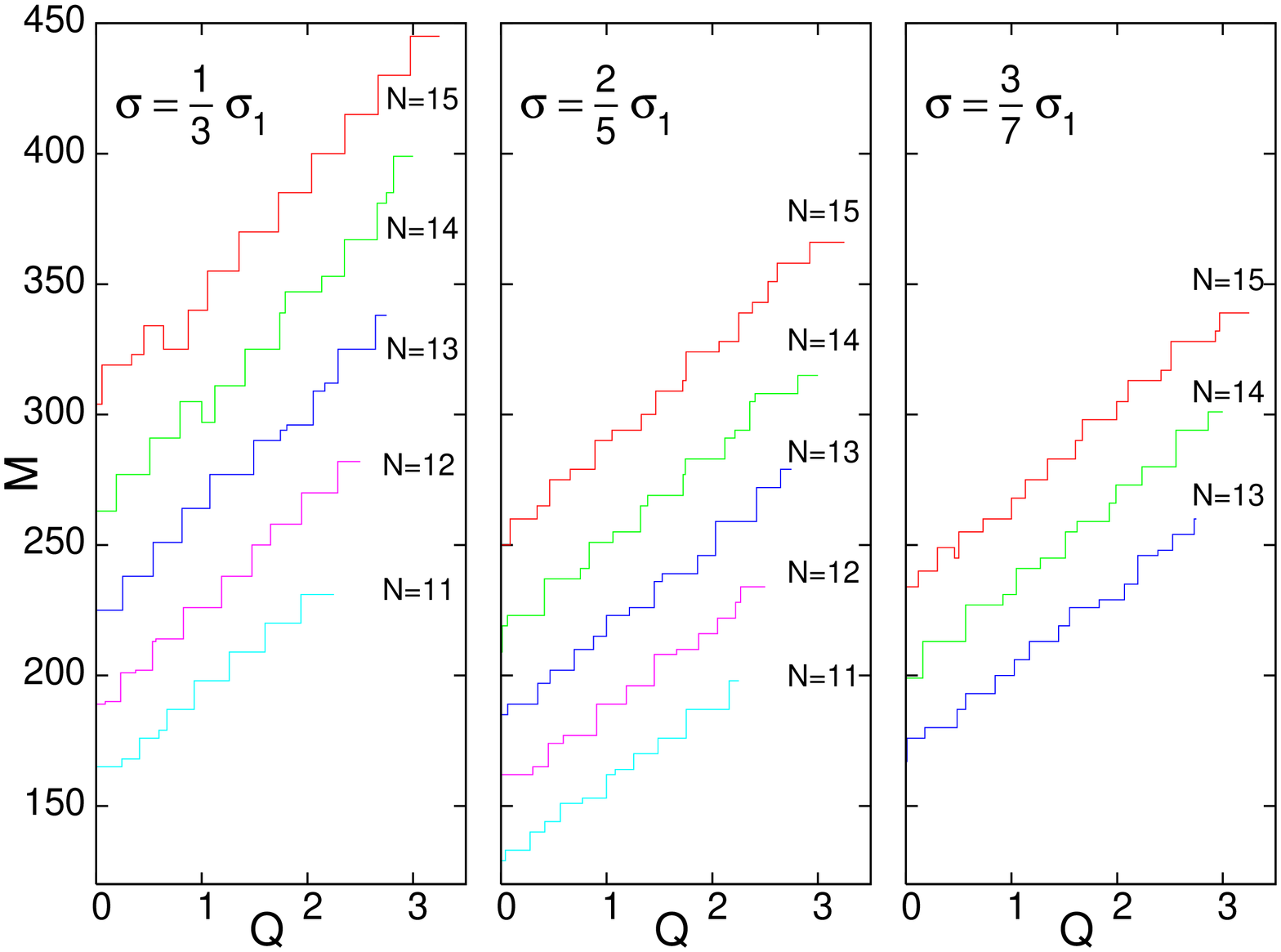,angle=-90,width=8.9cm}}

 {\small {\bf Fig.~3} Angular momentum $M(Q)$ of the ground state
changes in steps as the ``missing charge'' Q is tuned continuously.
The steps occur at the ground state degeneracies, when it costs no
energy to move a quasiparticle from the inner to the outer edge, and
can be associated with the resonant tunneling between the edges
through the bulk of the FQH state.}

 \vskip 0.1in

Figure 3 shows steps $M(Q)$ for $N$ from 11 through 15, as the
``missing charge'' Q is continuously increased.  The left panel shows
that three steps typically occur per $\Delta Q=1$ for all $N$.  The
general slope is consistent with Eq.~(\ref{mstarr1}), and most of the
steps in the left panel have $\Delta M=N$ precisely.  The data for the
$\frac{2}{5}$ fraction (the middle panel) shows steps occurring about
twice as often.  These observations are in line with our expectations
based on the quasi-periodicities seen in Fig.~1.  Careful examination
of the $\frac{2}{5}$ data shows that the change $M\rightarrow M+N$
occurs usually in two steps, although these steps are not always equal
to $N/2$.  The data for $\frac{3}{7}$ has been computed only for
$N\geq13$ as discussed above.  It shows behavior similar to the
$\frac{2}{5}$ data.  An expectation of 7 steps per $\Delta Q=1$ (or
$\Delta M=N/3$ as we would expect for $e^*=e/7$) cannot be confirmed.
This could mean that $N=15$ cluster size is not large enough to
distinguish the $\frac{3}{7}$ fraction.  It could also indicate a
genuine property of the $\frac{3}{7}$ fraction that needs to be
understood.

In conclusion, I have conducted a ``computer experiment'' designed to
model the key elements of the real experiment on resonant tunneling
through a quantum antidot.\cite{goldman-su} The data on small clusters
obtained using ED and CSR reveal a sequence of the ground state
reconstruction events consistent with the periodicity of the resonant
tunneling peaks observed experimentally.

The CSR approach employed here offers a number of advantages in
modeling many-body clusters.  In particular, truncation of the
single-particle basis, whenever possible (such as restricting the
range of $m$\cite{comment}) occurs automatically within CSR, which
discards irrelevant many-particle configurations and thus effectively
removes any single-particle state that contributes to none of the
relevant configurations.  This feature itself can be a major
simplification, because the relevance of a single-particle orbital
cannot be always judged in advance.

The term ``renormalization'' I use reflects the spirit of the
renormalization theory, yet I have not observed any defined fixed
point: the set of $K$ relevant configurations keeps changing slightly
upon convergence, with some marginally-relevant configurations being
replaced with other similarly relevant.

In general, performance of CSR may benefit if it is augmented with
rotations of the single-particle basis,\cite{dagotto98} though the
choice of the basis is fixed by symmetry in the case of the lowest
Landau level in disk geometry.

I would like to thank V.J. Goldman for numerous enlightening
conversations.  I also acknowledge stimulating discussions with
A.L. Aleiner.

\vskip -0.32in

\

\end{multicols}

\begin{references}
\bibitem{comment}
 The calculations were performed on a Pentium 4 workstation.  ED
results in Fig.~1 had single-particle angular momenta restricted to
empirically-found interval $(M-M^*+1)/N-1\leq m\leq (M-M^*-1)/N+3N-2$,
beyond which the occupation numbers are small.  This restriction
reduces the actual matrix size from $\sim10^9$ to about
$1.6\times10^7$.  In CSR calculations the above restriction on $m$ is
lifted, however I used $m\leq50$ for technical reasons.
\bibitem{comment1}
 The CSR calculation in Fig.~2 took 11 minutes.  A single point ED run
for the same cluster ($N=12$, $M=216$) took 24 hours.  The CPU time
comparison is meaningful only to a degree, because ED calculations are
disk intensive.
\bibitem{comment0}
 Technically, the basis expansion procedure is possible since the
Hamiltonian matrix is sparse and has a limited number of matrix
elements for each row (configuration).  I ``expand'' the
configurations one by one in the order of decreasing their weight.
For each configuration a number of new configurations related to it by
a matrix element is formed and added to the basis.  The process
continues until the target $K$ is achieved, and all the remaining
unexpanded configurations are discarded.  In order to enhance the
$K'/K$ ratio I limit the number of new configurations by considering
only the matrix elements with $\Delta m=\pm1$.  Different schemes for
choosing ``large'' matrix elements are possible and yield similar
results.
\end{references}
\end{document}